\documentstyle[11pt,axodraw]{article}
\begin{document}

\vskip 0.5cm
\centerline{\large\bf $B$ meson wave function in $k_T$ factorization}
\vskip 0.3cm
\centerline{Hsiang-nan Li$^{1,2}$ and Huei-Shih Liao$^1$}
\vskip 0.3cm
\centerline{$^1$Institute of Physics, Academia Sinica,
Taipei, Taiwan 115, Republic of China}
\vskip 0.3cm
\centerline{$^2$Department of Physics, National Cheng-Kung University,}\par
\centerline{Tainan, Taiwan 701, Republic of China}
\vskip 1.0cm
\centerline{\bf abstract}
\vskip 0.3cm

We study the asymptotic behavior of the $B$ meson wave function in
the framework of $k_T$ factorization theorem. We first construct a
definition of the $k_T$-dependent $B$ meson wave function, which
is free of light-cone divergences. Next-to-leading-order
corrections are then calculated based on this definition. The
treatment of different types of logarithms in the above corrections,
including the Sudakov logarithms, and those depending on a
renormalization scale and on an infrared regulator, is summarized.
The criticism raised in the literature on our resummation
formalism and Sudakov effect is responded. We show that the $B$
meson wave function remains normalizable after taking into account
renormalization-group evolution effects, contrary to the
observation derived in the collinear factorization theorem.

\vskip 1.0cm

\section{INTRODUCTION}

The $B$ meson distribution amplitude $\phi_+(k^+)$ plays an
essential role in a perturbative analysis of exclusive $B$ meson
decays based on collinear factorization theorem
\cite{BL,ER,CZS,CZ,L1}, where $k^+$ is the momentum carried by the
light spectator quark. Its behavior certainly matters, and has
been investigated in various approaches recently. Models of
$\phi_+$ with an exponential tail in the large $k^+$ region has
been proposed \cite{GN}. Neglecting three-parton distribution
amplitudes in a study by means of equations of motion
\cite{PB1,Braun:1990iv}, $\phi_+$ was found to be proportional to
a step function with a sharp drop at large $k^+$ \cite{KKQT}. The
asymptotic behavior of $\phi_+$ was also extracted from an
renormalization-group (RG) evolution equation derived in the
framework of collinear factorization theorem, which exhibits a
decrease slower than $1/k^+$ \cite{Neu03}. That is, the $B$ meson
distribution amplitude is not normalizable. This striking feature
has been confirmed in a QCD sum rule analysis \cite{BIK}, which
includes next-to-leading-order (NLO) perturbative corrections. For
a summary of the above progress, refer to \cite{LRev}. A similar
divergent normalization of the $B$ meson distribution function
involved in inclusive decays has been observed recently \cite{BM}.

A non-normalizable $B$ meson distribution amplitude does not cause
a problem in practice \cite{OL}. In a leading-order collinear
factorization formula, only the first inverse moment
$\lambda_B^{-1}(\mu)\equiv\int dk^+\phi_+(k^+)/k^+$ is relevant
\cite{BBNS,PK}, which is a convergent quantity. The factor $1/k^+$
comes from a hard kernel of a decay mode. Note that a hard kernel
would not be as simple as $1/k^+$ at higher orders, and
information of more moments is required. However, the
non-normalizability does introduce an ambiguity in defining the
$B$ meson decay constant $f_B$. The ambiguity can be understood
through the matrix element,
\begin{eqnarray}\label{LCDA}
   &&\langle\,0\,|\,\bar q(y)\,W_y(n_-)^{\dag}W_0(n_-)\,
\Gamma\,\not n_-\,h(0)\,
    |\bar B(v)\rangle
   = - \frac{iF(\mu)}{2}{\tilde \phi}_+(v\cdot y,\mu)\,
    \mbox{tr}\left( \Gamma\,\not n_-\,\frac{1+\not v}{2}\,\gamma_5
    \right)\;,
\end{eqnarray}
where the coordinate of the anti-quark field $\bar q$,
$y=(0,y^-,{\bf 0}_T)$, is parallel to the null vector
$n_-=(0,1,{\bf 0}_T)$, $h$ the rescaled $b$ quark field
characterized by the $B$ meson velocity $v$, $\mu$ the renormalization
scale, and $\Gamma$ represents a Dirac matrix. The factor
$W_y(n_-)$ denotes the Wilson line operator,
\begin{eqnarray}
\label{eq:WL.def}
W_y(n_-) = P \exp\left[-ig \int_0^\infty d\lambda
n_-\cdot A(y+\lambda n_-)\right]\;.
\end{eqnarray}
The quantity $F(\mu)$ is the HQET matrix element corresponding to
the asymptotic value of the product $f_B\sqrt{m_B}$ in the
heavy-quark limit. If the normalization $\widetilde\phi_+(v\cdot
y=0,\mu)$ is divergent, the definition of $f_B$ demands a further
arbitrary renormalization \cite{Np}.

We shall show that the above undesirable feature of $\tilde\phi_+$
is a consequence of adopting the collinear factorization theorem.
It has been known that the collinear factorization formulas of
many exclusive $B$ meson decays suffer end-point singularities
\cite{SHB}. We regard these singularities as an indication
\cite{Monr} that the $k_T$ factorization theorem
\cite{CCH,CE,LRS,BS,LS,NL2} is more appropriate for studying these
decays than the collinear factorization theorem. The perturbative
QCD (PQCD) approach \cite{LY1,CL,KLS,LUY} based on the $k_T$
factorization theorem has been developed. Retaining the parton
transverse momenta $k_T$ \cite{HS}, the end-point singularities
disappear \cite{TLS}, and the resultant predictions are in
agreement with most of experimental data \cite{KS02}. Viewing
these merits, it is very tempting to reanalyze the RG evolution
effect on the $B$ meson wave function (or the unintegrated $B$
meson distribution amplitude) in the $k_T$ factorization theorem.
Our conclusion is that the evolution effect does not drive the
asymptotic behavior of the $B$ meson wave function into $1/k^+$.
Therefore, the $B$ meson wave function is normalizable, and the
$B$ meson decay constant is well-defined.

In Sec.~II we find out a legitimate definition of the
$k_T$-dependent $B$ meson wave function. The NLO corrections to
the $B$ meson wave function are computed and compared to those in
\cite{Neu03} in Sec.~III. We respond to the criticism raised in
\cite{GS,LN03} on the PQCD formalism and on the Sudakov
effect in Sect.~IV. Section V is the conclusion.

\section{DEFINITIONS OF A WAVE FUNCTION}

We first construct the definition of the $B$ meson wave function
in the $k_T$ factorization theorem, which is nontrivial at all.
Hence, our formalism for the NLO calculation differs from those in
the literature \cite{MR,KPY}, which also involve the parton
transverse degrees of freedom. A gauge-invariant definition of the
$B$ meson wave functions ${\tilde \Phi}_+(v\cdot y, b, \mu)$ is
given via the nonlocal matrix element \cite{NL2},
\begin{eqnarray}
\langle 0|{\bar q}(y)W_y(n_-)^{\dag}I_{n_-;y,0} W_0(n_-)\not
n_-\Gamma h(0) |{\bar B}(v)\rangle\;,\label{nai}
\end{eqnarray}
as a naive extension of Eq.~(\ref{LCDA}) with $y=(0,y^-,{\bf b})$.
The two Wilson lines $W_y(n_-)$ and $W_0(n_-)$ must be connected
by a link $I_{n_-;y,0}$ at infinity in this case \cite{NL2,BJY}.

As pointed out in \cite{Co03}, Eq.~(\ref{nai}) contains additional
collinear divergences from the region with a loop momentum
parallel to $n_-$. These light-cone divergences, cancelling each
other as $b\to 0$, that is, as $\tilde\Phi_+(v\cdot y, b,
\mu)\to\widetilde\phi_+(v\cdot y, \mu)$, do not cause a problem in
the collinear factorization theorem. In the $k_T$ factorization
theorem, however, they must be subtracted in a gauge-invariant
way. Two modified definitions have been proposed in \cite{Co03}:
\begin{eqnarray}
& &\langle 0|{\bar q}(y) W_y(u)^{\dag}I_{u;y,0}W_0(u) \not
n_-\Gamma h(0)|{\bar B}(v)\rangle\;,\label{de1}\\
& &\frac{\langle 0|{\bar q}(y)
W_y(n_-)^{\dag}I_{n_-;y,0}W_0(n_-)\not n_-\Gamma h(0) |{\bar
B}(v)\rangle} {\langle
0|W_y(n_-)^{\dag}W_y(u')I_{n_-;y,0}I_{u';y,0}^{\dag}
W_0(n_-)W_0(u')^{\dag}|0\rangle}\;. \label{phim2}
\end{eqnarray}
In Eq.~(\ref{de1}) a non-light-like vector $u$ has been substituted
for the null vector $n_-$, so that no collinear divergence
is associated with the Wilson lines. In Eq.~(\ref{phim2}) $n_-$ is
maintained, but the light-cone divergences are regularized by the
denominator, which contains the same light-cone divergences as in
the numerator. As a gluon travels along $n_-$, it does not resolve
the detail of the valence quarks, which can then be replaced by
the Wilson lines in an arbitrary direction $u'$ \cite{Li96,LL}.
Both the above modifications with the appropriate Wilson links
are gauge-invariant. Nevertheless, the universality of the $B$
meson wave function is broken due to the appearance of the
auxiliary scale, for example, $\zeta=(k \cdot u)/\sqrt{u^2}$ from
Eq.~(\ref{de1}). Fortunately, the evolution in $\zeta$, the
so-called Sudakov evolution \cite{Co03}, can be derived using the
$k_T$-resummation technique \cite{Li96,CS}, such that the initial
condition of the evolution remains universal. Note that
Eqs.~(\ref{de1}) and (\ref{phim2}) do not approach the $B$ meson
distribution amplitude directly in the limit $b\to 0$ for general
$u$ and $u'$, but convolutions of hard kernels with the $B$ meson
distribution amplitude \cite{Co03,Li98}.

We have investigated the $O(\alpha_s)$ diagrams in Fig.~1
according to both modifications, and found that Eq.~(\ref{de1})
would change the ultraviolet structure of the quark-Wilson-line
vertex correction in Eq.~(\ref{nai}). This problem can be
explained using the pole term obtained from Fig.~1(c) (see
Eq.~(\ref{pe25}) in the Appendix),
\begin{eqnarray}
N^{(1)}_{c}\approx \frac{\alpha_sC_F}{4\pi}
\Gamma(\epsilon)\left[2-\left(\frac{4\zeta^2}{m_g^2}\right)^{-\epsilon}
\right]\;, \label{pe255}
\end{eqnarray}
$m_g$ being an infrared regulator. If taking the $u\to n_-$,.
i.e., $\zeta\to\infty$ limit in the above expression before making
the expansion in $\epsilon$, only the first term contributes to
the ultraviolet pole, which is $2/\epsilon$ in unit of
$\alpha_sC_F/(4\pi)$, the same as in Eq.~(\ref{nai}). If expanding
the factor $(4\zeta^2/m_g^2)^{-\epsilon}$ first (note that this
expansion makes sense for $u^2\not =0$, i.e., $\zeta\not
\to\infty$), the second term also contributes, and changes the
ultraviolet pole into $1/\epsilon$. Consequently, Eq.~(\ref{de1})
is governed by a RG evolution different from that of
Eq.~(\ref{nai}). In this work we shall adopt Eq.~(\ref{phim2}),
and demonstrate that the freedom in choosing the vector $u'$
allows a correct RG evolution of the $B$ meson wave function.

\section{$O(\alpha_s)$ CORRECTIONS}

The lowest-order evolution kernel for $\Phi_+(k^+,b,\mu)$ is
written as
\begin{eqnarray}
K^{(0)}(k^+,k^{\prime +},b,\mu)=\delta(k^+-k^{\prime +})\;,
\label{lo}
\end{eqnarray}
which implies that the light spectator quark, carrying only a
longitudinal momentum, is initially on-shell. It acquires the
transverse degrees of freedom through collinear gluon exchanges,
before participating a hard scattering \cite{NL2}. As indicated in
Eq.~(\ref{lo}), we perform $k_T$ factorization in the conjugate
$b$ space. We then calculate the $O(\alpha_s)$ corrections to
Eq.~(\ref{lo}) in dimensional regularization. A gluon mass $m_g$
is introduced to regularize the infrared divergences, so that we
can clearly distinguish the ultraviolet poles $1/\epsilon$ from
the infrared divergences represented by $\ln m_g$. As suggested in
\cite{Co03}, a small plus component is added to the null vector
$n_-$ in Eq.~(\ref{phim2}) at the intermediate step of the
calculation. That is, we start with the Wilson line in a
non-light-like direction $u$ for the numerator, and take the
$u\to n_-$ limit
eventually. Figures~1(a)-1(g) contribute to the numerator of
Eq.~(\ref{phim2}), and Figs.~1(a)-1(d) with the quark lines being
replaced by the Wilson lines along $u'$ contribute to the
denominator.

To highlight the difference between the collinear and $k_T$
factorizations, we present the loop integrals associated with
Figs.~1(a) and 1(b) and with Figs.~1(c) and 1(d),
\begin{eqnarray}
N^{(1)}_{a}+N^{(1)}_{b}&=&-ig^2C_F\mu^{2\epsilon}
\int\frac{d^{4-2\epsilon}l}{(2\pi)^{4-2\epsilon}} \frac{u\cdot
v}{(v\cdot l+i\epsilon)(l^2-m_g^2+i\epsilon)(u\cdot l+i\epsilon)} \nonumber\\
& &\times \left[\delta(k^+-k^{\prime +})-
\delta\left(k^+-k^{\prime +}+l^+\right) \exp(-i{\bf l}_T\cdot {\bf
b})\right]\;, \label{p2dBv}\\
N^{(1)}_{c}+N^{(1)}_{d}&=& \frac{i}{4}g^2C_F\mu^{2\epsilon}
\int\frac{d^{4-2\epsilon}l}{(2\pi)^{4-2\epsilon}}
tr\left[\gamma_\nu\frac{\not k'-\not l}{(k'-l)^2+i\epsilon}
\gamma_5\not n_- \not n_+\gamma_5\right]
\frac{1}{l^2-m_g^2+i\epsilon}\frac{u^\nu}{u\cdot
l+i\epsilon}\nonumber\\
& &\times\left[\delta(k^+-k^{\prime +})-\delta\left(k^+-k^{\prime
+}+l^+\right) \exp(-i{\bf l}_T\cdot {\bf b})\right]\;,
\label{p2eBrv}
\end{eqnarray}
respectively, where the arbitrary Dirac matrix $\Gamma$ has been
set to $\gamma_5$. In the above integrals we have made explicit
the $i\epsilon$ prescription in the propagators $1/v\cdot l$ and
$1/u\cdot l$, which follows the eikonal approximation of the
quark or gluon propagators the loop momentum $l$ flows through
\cite{BS}. Note the additional
Fourier factor $\exp(-i{\bf l}_T\cdot {\bf b})$ associated with
Figs.~1(b) and 1(d) \cite{BS,CS}. In the collinear factorization
theorem this Fourier factor disappears, corresponding to the $b\to
0$ limit. Moreover, for $u=n_-$, we obtain, from
Eqs.~(\ref{p2dBv}) and (\ref{p2eBrv}), the counterterm identical
to that found in the collinear factorization theorem \cite{Neu03},
\begin{eqnarray}
-\frac{\alpha_sC_F}{4\pi}\frac{2}{\epsilon}\left[
\frac{k^+}{k^{\prime +}}\frac{\theta(k^{\prime +}-k^+)}{(k^{\prime
+}-k^+)_+} +\frac{\theta(k^+-k^{\prime +})}{(k^+-k^{\prime
+})_+}\right]\;. \label{peB3}
\end{eqnarray}
The resultant anomalous dimension contributes to the splitting
kernel in the RG evolution equation, that determines the
asymptotic behavior of the $B$ meson distribution amplitude
\cite{Neu03}. The ultraviolet pole $1/\epsilon$ arises from the
integration over the transverse loop momentum $l_T$ up to infinity.
It is then expected that Eq.~(\ref{peB3}) will be absent in the $k_T$
factorization theorem due to the suppression in the large $l_T$
region from $\exp(-i{\bf l}_T\cdot {\bf b})$.


The $O(\alpha_s)$ corrections $N^{(1)}_i(k^+,k^{\prime +},b,\mu)$
from Figs.~1(i), $i=a\cdots e$, to the numerator of
Eq.~(\ref{phim2}) are summarized below:
\begin{eqnarray}
N^{(1)}_{a}&=&-\frac{\alpha_sC_F}{4\pi}
\ln(2\nu^2)\left(\frac{1}{\epsilon}
+\ln\frac{4\pi\mu^2}{m_g^2e^{\gamma_E}}\right)\delta(k^+-k^{\prime
+})\;, \label{z1d}\\
N^{(1)}_b&=&-\frac{\alpha_sC_F}{4\pi}\left[\ln(2\nu^2)
\ln\frac{m_g^2 b^2e^{2\gamma_E}}{4}\delta(k^+-k^{\prime +})\right.
\nonumber\\
& &\left.+4\frac{\theta(k^{\prime +}-k^+)}{(k^{\prime +}-k^+)_+}
K_0\left(2\nu(k^{\prime
+}-k^+)b\right)-4\frac{\theta(k^+-k^{\prime +})}{(k^+-k^{\prime
+})_+} K_0\left(\sqrt{2}(k^+-k^{\prime +})b\right) \right]\;,
\label{pde}\\
N^{(1)}_c&=&\frac{\alpha_sC_F}{4\pi}
\left[\frac{1}{\epsilon}-2\ln^2\frac{2\nu k^+}{m_g} +2
\ln\frac{2\nu k^+}{m_g}+\ln\frac{4\pi\mu^2}{m_g^2e^{\gamma_E}}
-\frac{5}{6}\pi^2\right] \delta(k^+-k^{\prime +})\;,
\label{peB1v}\\
N^{(1)}_d&=&-\frac{\alpha_sC_F}{4\pi}\Bigg\{\left[
\ln\frac{2\nu^2k^{+2} b
e^{\gamma_E}}{m_g}\ln\frac{m_g^2b^2e^{2\gamma_E}}{4}
+\frac{\pi^2}{3}\right]\delta(k^+-k^{\prime +}) \nonumber\\
& &-4\frac{k^+\theta(k^{\prime +}-k^+)}{k^{\prime +}(k^{\prime
+}-k^+)_+} \left[K_0\left(\sqrt{k^+/k^{\prime +}}m_gb\right)
-K_0\left(2\nu(k^{\prime +}-k^+)b\right)\right] \Bigg\}\;,
\label{peB1r}\\
N^{(1)}_e&=&\frac{\alpha_sC_F}{4\pi}\Bigg\{\left[\ln\frac{k^{+2} b
e^{\gamma_E}}{m_g}\ln\frac{m_g^2b^2e^{2\gamma_E}}{4}
+\frac{\pi^2}{3}\right]\delta(k^+-k^{\prime +}) \nonumber\\
& &-4\frac{k^+\theta(k^{\prime +}-k^+)}{k^{\prime +}(k^{\prime
+}-k^+)_+} \left[K_0\left(\sqrt{k^+/k^{\prime +}}m_gb\right)
-K_0\left(\sqrt{2}(k^{\prime +}-k^+)b\right)\right] \Bigg\}\;.
\label{p2bB1}
\end{eqnarray}
For their detailed derivation, refer to the Appendix. The
auxiliary parameter $\nu=(n_+ \cdot u)/\sqrt{u^2}$, defined via
$\zeta\equiv \nu k^+$, denotes the $u$ dependence, $\gamma_E$ is
the Euler constant, and the subscript ``+" in the factor
$1/(k^{\prime +}-k^+)_+$ represents the ``plus" distribution.  The
self-energy corrections to the heavy quark field $h$ and to the
light spectator quark $\bar q$ are
\begin{eqnarray}
N^{(1)}_f &=&\frac{\alpha_s C_F}{4\pi}\left(\frac{1}{\epsilon}
+\ln\frac{4\pi\mu^2}{m_g^2 e^{\gamma_E}}\right)
\delta(k^+-k^{\prime +})\;,
\label{psia1}\\
N^{(1)}_g&=& -\frac{\alpha_s C_F}{4\pi}\left(\frac{1}{2\epsilon}
+\frac{1}{2}\ln\frac{4\pi\mu^2}{m_g^2
e^{\gamma_E}}-\frac{1}{4}\right) \delta(k^+-k^{\prime +})\;.
\label{psic1}
\end{eqnarray}


Some remarks are in order. The logarithms $\ln(2\nu^2)$ denote
the light-cone collinear divergences mentioned before. Equation
(\ref{z1d}) does not contain a double pole $1/\epsilon^2$ observed
in \cite{Neu03} due to the replacement of the null vector $n_-$
by the non-light-like vector $u$. For a similar reason, the
single-pole term $\ln(\mu/k^+)/\epsilon$, which leads to the type
of Sudakov logarithms in the collinear factorization theorem
\cite{Neu03}, does not exist. As expected, Eq.~(\ref{pde}), with
the suppression from the Fourier factor, does not generate the
ultraviolet pole in Eq.~(\ref{peB3}). Because of the Bessel
function $K_0$, the splitting effect from the plus distribution is
negligible in the asymptotic region with large $k^+$. Equation
(\ref{peB1v}) produces the double logarithm $\ln^2(k^+/m_g)$,
which is not yet in the form of Sudakov logarithms in the $k_T$
factorization theorem. After combining Eqs.~(\ref{peB1v}) and
(\ref{peB1r}), we derive the standard $k_T$-dependent
infrared-finite Sudakov logarithms,
\begin{eqnarray}
-\frac{\alpha_sC_F}{2\pi}\left[\ln^2(k^+b) - (1-2\gamma_E)\ln(k^+
b)\right]\;,\label{ss}
\end{eqnarray}
with the first (second) term being leading (next-to-leading). The
$\nu$ dependence is not made explicit, since it will be cancelled
by that from the denominator. These logarithms should be resummed
to all orders using the technique in \cite{LY1}. The double
infrared logarithm $\ln^2 m_g$ in Eq.~(\ref{p2bB1}) is new, which
does not exist in radiative corrections to a light meson process
\cite{MR}. The important splitting effects, proportional to
$K_0(\sqrt{k^+/k^{\prime +}}m_gb)\approx \ln(m_g b)$, cancel
between Eqs.~(\ref{peB1r}) and (\ref{p2bB1}).

The $O(\alpha_s)$ corrections to the denominator are computed in a
similar way, but with $\delta\left(k^+-k^{\prime +}\right)$ being
substituted for $\delta\left(k^+-k^{\prime +}+l^+\right)$. This
substitution is made in that the denominator is to remove the
light-cone divergences arising from $l^+\to 0$. Hence, the
splitting terms, i.e., the plus distributions in
Eqs.~(\ref{z1d})-(\ref{p2bB1}), disappear. We choose $u'=v$ for the
incoming Wilson line (along the $b$ quark), and a different $u'$
for the outgoing Wilson line, such that the ultraviolet structure
of the quark-Wilson-line vertex correction the same as in
Eq.~(\ref{nai}) is recovered. We emphasize that other choices of
$u'$ are equivalent, in view that the resultant $B$ meson wave
functions all collect the same soft structure of an exclusive
decay. The expressions are summarized below:
\begin{eqnarray}
D^{(1)}_{a}&=&N^{(1)}_{a}\;, \\
D^{(1)}_{b}&=&-\frac{\alpha_sC_F}{4\pi}\ln(2\nu^2)
\ln\frac{m_g^2 b^2e^{2\gamma_E}}{4}\delta(k^+-k^{\prime +})
\;,\\
D^{(1)}_{c}&=&-\frac{\alpha_sC_F}{4\pi}
\ln(4\nu^2\nu^{'2})\left(\frac{1}{\epsilon}
+\ln\frac{4\pi\mu^2}{m_g^2e^{\gamma_E}}\right)\delta(k^+-k^{\prime
+})\;, \label{da}\\
D^{(1)}_d&=&-\frac{\alpha_sC_F}{4\pi} \ln(4\nu^2\nu^{'2})
\ln\frac{m_g^2 b^2e^{2\gamma_E}}{4}\delta(k^+-k^{\prime +})\;,
\label{db}
\end{eqnarray}
with the auxiliary parameter $\nu'=(u'\cdot n_-)/\sqrt{u^{'2}}$.
It is easy to check that the sum of the above corrections is free
of the infrared cutoff $m_g$. That is, the denominator in
Eq.~(\ref{phim2}) does not alter the soft structure of the
numerator, i.e., of Eq.~(\ref{nai}), as requested above. According
to our prescription, we set $\ln(4\nu^{'2})$ to unity.


The total one-loop correction $K^{(1)}$ to Eq.~(\ref{lo}) is then
written as,
\begin{eqnarray}
K^{(1)}&=&\sum_{j=a}^{g}N^{(1)}_j-\sum_{j=a}^{d}D^{(1)}_j\;,
\nonumber\\
&=&\frac{\alpha_s
C_F}{4\pi}\Bigg\{\left(\frac{5}{2}+\ln\nu^2\right)\left[
\frac{1}{\epsilon} +\ln\left(\pi
e^{\gamma_E}\mu^2b^2\right)\right]-2\ln^2(\nu k^+ b) +
2(1-2\gamma_E)\ln(\nu k^+ b)\nonumber\\
& & -(5-2\gamma_E)\ln\left(\frac{m_g b}{2}\right)+2\ln\frac{k^{+2}
b}{m_g}\ln\frac{m_gbe^{\gamma_E}}{2} +\frac{1}{4}-\frac{5}{6}\pi^2
-3\gamma_E\Bigg\}\;. \label{tot}
\end{eqnarray}
The ultraviolet pole $5/(2\epsilon)$ in unit of
$\alpha_sC_F/(4\pi)$ is the same as the corresponding one derived
in Eq.~(8) of \cite{Neu03} under our prescription for fixing $u'$.
Note that it differs from $3/\epsilon$ in \cite{LY1}, since it is
the $b$ quark field, instead of the rescaled one, that was adopted
to define the $B$ meson wave function in \cite{LY1}. The pole
$5/(2\epsilon)$ should be partitioned in the way that
$3/(2\epsilon)$ contributes to the factor $F(\mu)$ in
Eq.~(\ref{LCDA}) and $1/\epsilon$ to $\Phi_+(k^+,b,\mu)$. Here we
do not perform such a partition. The splitting terms, which either
vanish in the asymptotic region with large $k^+$ or cancel between
Eqs.~(\ref{peB1r}) and (\ref{p2bB1}), have been dropped.

The treatment of each term in Eq.~(\ref{tot}) is explained as
follows. The ultraviolet pole together with the constants are
subtracted in a renormalization procedure. The logarithm $\ln(\mu
b)$ is then summed to all orders using a standard RG evolution
equation \cite{Neu03}, giving an exponential $R(b,\mu,\nu)$. The
Sudakov logarithms $\alpha_s\ln^2(\nu k^+ b)$ and $\alpha_s\ln(\nu
k^+ b)$ are resummed, leading to the Sudakov factor $S(k^+,b,\nu)$
\cite{LY1}. The evolution of the $B$ meson wave function from
Eq.~(\ref{phim2}) is then given by
\begin{eqnarray}
\Phi_+(k^+,b,\mu)=S(k^+,b,\nu)\phi_+(k^+,b,\mu)\;,\;\;\;\;
\phi_+(k^+,b,\mu)=R(b,\mu,\nu)\phi_+(k^+,b,\mu=1/b)\;.
\label{rphi}
\end{eqnarray}
The logarithms $\ln(m_gb)$ and $\ln(k^{+2}b/m_g)\ln(m_gb)$,
representing the soft structure of the $B$ meson wave function,
are absorbed into the initial condition $\phi_+(k^+,b,\mu=1/b)$ of
the above evolution. They are then used to subtract the infrared
divergences in the evaluation of hard kernels, i.e., in the
so-called ``matching" procedure.

The exponentials in Eq.~(\ref{rphi}) are quoted from
\cite{LY1,Li96} as
\begin{eqnarray}
S(k^+,b,\nu)&=&\exp\Bigg\{-\int_{1/b}^{k^+}\frac{d\bar\mu}
{\bar\mu}\left[\ln\frac{k^+}{\bar\mu}A(\alpha_s(\bar\mu))
+B(\nu,\alpha_s(\bar\mu))\right]\Bigg\}\;, \label{sud}\\
R(b,\mu,\nu)&=&\exp\left[-\int_{1/b}^{\mu}
\frac{d\bar{\mu}}{\bar{\mu}}\gamma (\alpha_s(\bar{\mu}))\right]\;,
\label{r}
\end{eqnarray}
with the one-loop anomalous dimensions,
\begin{eqnarray}
A&=&\frac{\alpha_s}{\pi}C_F\;,
\label{a} \\
B&=&\frac{\alpha_s}{2\pi}C_F
\ln\left(\nu^2e^{2\gamma_E-1}\right)\;,
\label{b}\\
\gamma&=&-\frac{\alpha_s}{4\pi}C_F\left(5+2\ln\nu^2\right)\;,
\end{eqnarray}
where the running of the coupling constant $\alpha_s$ has been
taken into account. It can be confirmed trivially that the exponent
of Eq.~(\ref{sud}) is identical to the Sudakov logarithms in
Eq.~(\ref{tot}), if the running of $\alpha_s$ is frozen. In a
practical analysis, the scale $\mu$ is set to a hard scale, which
is usually of order $k^+$. The $\nu$-dependences then cancel
between $S(k^+,b,\nu)$ and $R(b,\mu=k^+,\nu)$, such that the $B$
meson wave function $\Phi_+(k^+,b,\mu=k^+)$ does not depend $\nu$.
This cancellation is equivalent to that of the light-cone
divergences, in agreement with the speculation in \cite{Co03}.
After the above cancellation, the Sudakov exponent in
Eq.~(\ref{sud}) reduces to the logarithms in Eq.~(\ref{ss}), if
neglecting the running of $\alpha_s$, and the anomalous dimension
$\gamma$ is equal to $-5$ in unit of $\alpha_s/(4\pi)$. We then
bring the Wilson line direction $u$ back to the null vector $n_-$
as stated before.

At last, we discuss the normalization of the $B$ meson wave
function $\Phi_+(k^+,b,\mu)$ in the $k_T$ factorization theorem,
which is defined as
\begin{eqnarray}
\int_0^{\infty}dk^+\lim_{b\to 1/k^+}\Phi_+(k^+,b,\mu)=
\int_0^{\infty}dk^+\phi_+(k^+,b=1/k^+,\mu)\;. \label{rp2}
\end{eqnarray}
The Sudakov factor in Eq.~(\ref{rphi}) becomes identity in the
limit $b\to 1/k^+$, which approaches zero in the heavy quark limit
for a fixed momentum fraction $x\equiv k^+/(m_Bv^+)$. In the above
limit the splitting terms proportional to the Bessel function
$K_0$ remain finite, and will not contribute to the evolution
kernel. It is then obvious from Eq.~(\ref{rphi}) that the
normalizability of the $B$ meson distribution amplitude is not
spoiled by the RG evolution effect, when evaluated according to
Eq.~(\ref{rp2}).

\section{RESPONSE TO THE CRITICISM}

After completing the NLO calculation for the $B$ meson wave
function in the $k_T$ factorization theorem, we are ready to
respond to the cricitism raised by Descotes-Genon and Sachrajda
\cite{GS} and by Lange and Neubert \cite{LN03}, which concerned
the PQCD formalism and the Sudakov effect. We fully
recognize that it is not easy to appreciate the delicacies of
different approaches, and that misunderstandings are unavoidable.
We hope that this section helps clarify these misunderstandings.

It was concluded that a heavy-to-light form factor is not
calculable in practice due to the ignorance of the heavy meson
wave functions \cite{GS}. The word ``calculable" is perhaps
confusing. It is more appropriate to use ``factorizable", which
means that a physical quantity can be written to all orders of
$\alpha_s$ as a factorization formula containing a hard kernel
(Wilson coefficient) and wave functions, plus jet functions,
Sudakov factors $\cdots$. Then a form factor is factorizable in
the PQCD approach based on the $k_T$ factorization theorem because
of the absence of the end-point singularities. A form factor is
not factorizable in QCD-improved factorization (QCDF) (considering
only the leading contribution) \cite{BBNS}, and partially
factorizable in soft-collinear effective theory (SCET) \cite{BPS},
since both factorizable and nonfactorizable pieces exist at
leading level. Therefore, a heavy meson wave function plays the
role of an input in the PQCD approach the same as the form factor
does in QCDF, and is determined by the value of the form factor
from experimental data, lattice QCD, or sum rules. The
inappropriate criticism in \cite{GS} is thus a conceptual
misunderstanding of the PQCD approach.

A difference between the pion form factor and the $B\to\pi$
transition form factor was pointed out \cite{GS}: the former does
not contain an end-point singularity in the collinear
factorization theorem, but the latter does. Hence, it was
questioned whether the PQCD approach, working for the former, can
be extended to the latter. Our opinion is that both collinear and
$k_T$ factorization theorems are applicable to the pion form
factor or to the decay $B\to\gamma l\nu$ \cite{DS}, and the
numerical outcomes are not very different, because there is no
end-point singularity. However, the end-point singularity in the
collinear factorization formula of the $B\to\pi$ form factor
demands the use of the $k_T$ factorization theorem, which is more
conservative than the collinear one: the parton transverse momenta
should not be treated as a pure higher-power effect, when the
end-point region of a parton momentum fraction is important. This
is the motivation to develop the PQCD approach, and it makes sense
to compare its predictions for two-body nonleptonic $B$ meson
decays with experimental data.

It was stated that the $b$ quark line a collinear gluon attaches
could not be approximated by a Wilson line in the direction $v$,
that is, could not be replaced by the rescaled $b$ quark line
\cite{GS}. In fact, the approximation holds, no matter a soft or
collinear gluon attaches the $b$ quark, in that it does not change
the soft or collinear divergence of a loop integral. This is
exactly the idea to remove the light-cone divergences in the
numerator of Eq.~(\ref{phim2}) by introducing the denominator,
where a Wilson line is substituted for the $b$ quark.

It has been argued that Fig.~1(e) does not contain the double
logarithm $\ln^2 (k\cdot v/\sqrt{v^2})$, and the invariant
$(k\cdot v)^2/v^2$ is irrelevant in the Sudakov resummation
\cite{LY1}. This argument was doubted in \cite{GS}. The explicit
$O(\alpha_s)$ result in Eq.~(\ref{p2bB1}) has clarified the issue:
$\zeta$ is the only relevant invariant. It is well-known that a
double logarithm arises from a vertex correction \cite{KR}, such
as Fig.~1(c), instead of from the type of corrections like
Fig.~1(e).

An expression equivalent to Eq.~(\ref{sud}) was given by \cite{GS}
\begin{equation}
S(k^+,b,\nu)=\exp\Bigg\{-\int_{C_1/b}^{C_2\nu k^+} \frac{d
\bar\mu}{\bar\mu} \left[\ln\left(\frac{C_2\nu k^+}{\bar\mu}
\right)A(C_1,\alpha_s(\bar\mu))+B(C_1,C_2,\alpha_s(\bar\mu))\right]\Bigg\}\;,
\label{fsl1}
\end{equation}
where $A(C_1,\alpha_s)$ is equal to $A(\alpha_s)$ in Eq.~(\ref{a})
at one-loop level, and
\begin{eqnarray}
B(C_1,C_2,\alpha_s)=\frac{\alpha_s}{2\pi}C_F
\ln\left(\frac{e^{2\gamma_E-1}C_1^2} {C_2^2}\right)\;.
\end{eqnarray}
Our choice $C_1=C_2=1$ \cite{LY1,KLS}, questioned in \cite{GS}, is
now justified, since it indeed reproduces the logarithms in
Eq.~(\ref{tot}). Other choices of $C_1$ and $C_2$ are equally
fine: they lead to a change only in the next-to-leading
logarithms, which can be compensated by the corresponding change
in hard kernels. Hence, the choice of $C_1$ and $C_2$ is not a
problem from the viewpoint of factorization theorem.

It was claimed that the wave function in Eq.~(\ref{de1}) defined
in terms of the non-light-like vector $u$ could not be related to
the standard definition in terms of the null vector $n_-$
\cite{GS}. As explained in Sec.~II, the $\nu$ dependence can be
grouped into the Sudakov factor, and the initial condition of the
Sudakov evolution is identified as the standard definition. The
validity of Eq.~(\ref{rphi}) in the large $b$ region was also
challenged, because it suffers a large $O(\alpha_s(1/b))$
correction. The treatment of this correction, not
multiplied by a logarithm, is a matter of factorization scheme: it
corresponds to the constant terms in Eq.~(\ref{tot}), and is
allowed to shift freely between a wave function and a hard kernel.
This shift is similar to that of the next-to-leading Sudakov
logarithms resulted in by varying the parameters $C_{1,2}$ in
Eq.~(\ref{fsl1}). Therefore, it is always possible to choose a
factorization scheme for a NLO evaluation of a hard kernel, such
that Eq.~(\ref{rphi}) holds.

The explicit $\nu$ dependence of the Sudakov factor derived from
Eq.~(\ref{de1}) \cite{LY1} was pointed out in \cite{GS}. It has
been known that this $\nu$ dependence is cancelled by that of a
soft function, which collects irreducible soft gluons to all
orders. The cancellation has been demonstrated in the processes
including Landshoff scattering \cite{BS}, deep inelastic
scattering \cite{Li96}, Drell-Yan production \cite{Li96},
inclusive semileptonic $B$ meson decays \cite{Li96}, dijet
production \cite{KOS}, and the $B\to D\pi$ decays \cite{LT97}. We
believe that such a cancellation is general, though having not yet
explored all other processes, since a physical quantity should not
depend on this artificial dependence.

There are two leading-twist $B$ meson wave functions $\Phi_+$ and
$\Phi_-$ \cite{GN}, the latter being defined by, for example, the
matrix element in Eq.~(\ref{phim2}) with $\not n_-$ replaced by
$\not n_+$, where $n_+=(1,0,{\bf 0}_T)$ is another null vector. It
was claimed \cite{GS} that the equality $\Phi_+=\Phi_-$ was
assumed in the PQCD approach \cite{KLS}. We make clear that this
equality was never postulated. Precisely speaking, the $B$ meson
wave function $\Phi_B$ adopted in \cite{KLS} is identified as
$\Phi_+$ discussed here, and another wave function $\bar\Phi_B$,
appearing as the combination $\Phi_+-\Phi_-$, is numerically
negligible as confirmed in \cite{WY}. The above combination was
found to be important, when its contribution to a single term in
the full expression of the form factor was investigated \cite{GS}.
An observation based on such an incomplete analysis is certainly
not solid. Because the wave function $\Phi_-$ does not appear in the
leading PQCD formalism, it is not urgent to discuss the
corresponding Sudakov resummation.

It was concluded \cite{LN03} that the Sudakov effect is not
important for the soft contribution to the $B\to\pi$ form factor.
First, we emphasize that the Sudakov logarithms studied in SCET
\cite{LN03} differ from what we discussed here and adopted in the
PQCD approach \cite{KLS}: the former appear in the Wilson
coefficient associated with the soft form factor in the collinear
factorization theorem, while the latter come from the wave
functions in the $k_T$ factorization theorem. The difference
manifests itself in the explicit expressions of the Sudakov
factors: the latter is $k_T$-dependent, but the former is not.
Second, there is no conflict between the conclusions in \cite{KLS}
and in \cite{LN03}. The weak Sudakov suppression in \cite{LN03}
refers to that on the whole form factor. The strong suppression in
PQCD applies only to the end-point region of a momentum fraction
(a form factor is factorizable in PQCD), and the suppression away
from the end point is weak. Note that the strong Sudakov effect
has been confirmed in \cite{GS} (see Page 271) for the model of
the $B$ meson wave function proposed in \cite{KLS,LM99}. Speaking
of the whole form factor, whose contribution mainly arises from
the non-end-point region, the suppression studied in PQCD is not
significant either. For a more detailed explanation on this issue,
refer to \cite{KKLL}.

\section{CONCLUSION}

In this paper we have surveyed the definitions of a wave function
in the $k_T$ factorization theorem given in
Eqs.~(\ref{nai})-(\ref{phim2}). The naive one in Eq.~(\ref{nai})
contains additional light-cone collinear divergences, which cancel
in a distribution amplitude in the collinear factorization
theorem. These light-cone divergences are removed in the modified
definitions of Eqs.~(\ref{de1}) and (\ref{phim2}) in
a gauge-invariant way. However, Eq.~(\ref{de1}), in which the
Wilson line has been rotated away from the light cone to an arbitrary
direction $u$, alters the ultraviolet structure of Eq.~(\ref{nai}).
Certainly, this change is not a problem, similar to the fact that
the ultraviolet structure of a heavy-light currnt is changed under
the HQET approximation. All the definitions of the $B$ meson wave
function are equivalent, as long as they collect the same soft
structure of an exclusive decay. We have found that it is possible to
maintain the ultraviolet structure by adopting Eq.~(\ref{phim2}).
The dependence on a general $u$ or $u'$ can
be factored into the Sudakov factor, such that the wave function, as
the initial condition of the Sudakov evolution, is
gauge-invariant and universal \cite{Co03}. Eventually,
the $u$ or $u'$ dependence of the Sudakov factor will be cancelled
by that of a soft function as making predictions for a physical
quantity.

We have calculated the $O(\alpha_s)$ corrections from
Figs.~1(a)-1(g) to the $B$ meson wave function following the
definition in Eq.~(\ref{phim2}), which contain three types of
logarithms $\ln(k^+b)$, $\ln(\mu b)$ and $\ln(m_gb)$. The leading
and next-to-leading infrared-finite Sudakov logarithms $\ln(k^+b)$
have been verified, which are consistent with
the Sudakov exponent adopted in our previous works. It has been
observed that Figs.~1(b) and 1(d) do not generate the ultraviolet
poles from the integration over the transverse loop momenta $l_T$
due to the suppression from the Fourier factor $\exp(-i{\bf
l}_T\cdot {\bf b})$. Hence, the RG evolution from the summation
of $\ln(\mu b)$ is trivial. We have explained that the small $b$ limit
should be taken as $b\to 1/k^+$ in the $k_T$ factorization
theorem, under which the Sudakov evolution factor becomes
identity, and Figs.~1(b), 1(d) and 1(e) remain ultraviolet finite.
As a consequence, the RG evolution effect does not spoil the
normalizability of the $B$ meson wave function, when evaluated
according to Eq.~(\ref{rp2}). This is another indication that the
$k_T$ factorization theorem is a more appropriate framework for
studying exclusive $B$ meson decays than the collinear
factorization theorem. At last, the infrared logarithms $\ln(m_gb)$
are used to subtract the corresponding infrared divergences in
the computation of hard kernels.

Our NLO calculation is similar to that performed in \cite{MR},
where the conjugate $b$ space was also introduced. However, it was
the $\gamma^*\gamma\to\pi^0$ amplitude, instead of the pion wave
function, that was studied in \cite{MR}. Therefore, the issues of
the undesirable light-cone collinear divergences and of a
legitimate definition of a $k_T$-dependent wave function were not
addressed. The $O(\alpha_s)$ corrections to the $B\to\gamma l \nu$
decay amplitude were computed in a different way in \cite{KPY}.
First, the issues mentioned above were not addressed either.
Second, a parton was assumed to carry a transverse momentum
initially, and the conjugate $b$ space was not introduced. Third,
a different type of double logarithms $\ln^2(m_B/k^+)$ was
resummed, leading to the so-called threshold resummation
\cite{L5}. Our opinion for proceeding a NLO analysis in the $k_T$
factorization theorem is that one must define a valid
$k_T$-dependent wave function first (under a factorization scheme
as stated in Sec.~IV). Next, one computes the $O(\alpha_s)$
corrections to the full parton-level amplitude, from which the
wave function also evaluated at $O(\alpha_s)$ is subtracted. This
subtraction results in an infrared-finite hard kernel,
which is then substituted into a $k_T$ factorization formula to estimate
the NLO effect. Therefore, our work provides a basis of the above
systematic procedure.

\vskip 0.5cm We thank C.K. Chua, J. Collins, S. Gardner, B. Melic, Y.Y.
Keum, T. Kurimoto, C.D. Lu, M. Neubert, E.A. Paschos, S. Recksiegel,
and A.I. Sanda for helpful discussions. This work was supported in
part by the National Science Council of R.O.C. under Grant No.
NSC-92-2112-M-001-030 and by Taipei branch of the National Center
for Theoretical Sciences of R.O.C..

\appendix

\section{DETAIL OF CALCULATION}

We present the details of the $O(\alpha_s)$ calculation in this
Appendix. We assume the small plus component added to $n_-$ to be
negative, $u^+<0$, for convenience. One can always work out a loop
integral, whose result is a function of $u^2$, in the $u^2 < 0$
($u^2> 0$) region, and then analytically continue the result into
the $u^2> 0$ ($u^2 < 0$) region. Applying contour integration in
the light-cone coordinates, we obtain the integral for the
numerator of Eq.~(\ref{phim2}) associated with Fig.~1(a),
\begin{eqnarray}
N^{(1)}_{a}&=&ig^2C_F\mu^{2\epsilon} \frac{2\pi i u\cdot
v}{u^+v^--u^-v^+}
\int\frac{d^{2-2\epsilon}l_T}{(2\pi)^{4-2\epsilon}}
\Bigg\{\int_0^\infty dl^+\frac{u^+}{l^+[2u^-l^{+2}+
u^+(l_T^2+m_g^2)]}
\nonumber\\
& &+\int_{-\infty}^0dl^+\frac{v^+}{l^+[2v^-l^{+2}+
v^+(l_T^2+m_g^2)]}\Bigg\}\delta(k^+-k^{\prime +})\;, \label{ad1}
\end{eqnarray}
where the first term corresponds to the pole $l^-=-u^-l^+/u^++i\epsilon$
for $l^+>0$, and the second term to the pole $l^-=-v^-l^+/v^+-i\epsilon$
for $l^+<0$. The integration over $l_T$ leads to
\begin{eqnarray}
N^{(1)}_{a}&=&-\frac{\alpha_sC_F}{2\pi} \frac{u\cdot
v}{u^-v^+-u^+v^-}\left(\frac{4\pi\mu^2}{m_g^2}\right)^\epsilon
\Gamma(\epsilon)
\nonumber\\
& &\times\int_0^\infty\frac{dt}{t^{1-2\delta}}\left[
\left(\frac{2v^-}{v^+}t^{2}+1\right)^{-\epsilon}
-\left(\frac{2u^-}{u^+}t^{2}+1\right)^{-\epsilon}
\right]\delta(k^+-k^{\prime +})\;, \label{ad11}
\end{eqnarray}
where the variable change $l^+=m_gt$ has been applied. It is easy
to observe that the soft divergences in the above two terms
cancel. Hence, the small parameter $\delta$, introduced for
convenience, will approach zero eventually. Working out the
integration over $t$, we have
\begin{eqnarray}
N^{(1)}_{a}&=&-\frac{\alpha_sC_F}{4\pi}\frac{u\cdot
v}{\sqrt{(u\cdot
v)^2-u^2v^2}}\left(\frac{4\pi\mu^2}{m_g^2}\right)^\epsilon
\Gamma(\epsilon)B(\delta,\epsilon-\delta)\left[
\left(\frac{v^+}{2v^-}\right)^\delta
-\left(\frac{u^+}{2u^-}\right)^\delta\right] \delta(k^+-k^{\prime
+})\;,
\end{eqnarray}
whose $\delta\to 0$ limit is given by
\begin{eqnarray}
N^{(1)}_{a}&=&-\frac{\alpha_sC_F}{4\pi}\frac{u\cdot
v}{\sqrt{(u\cdot
v)^2-u^2v^2}}\left(\frac{4\pi\mu^2}{m_g^2}\right)^\epsilon
\Gamma(\epsilon)\ln\frac{u^-v^+}{u^+v^-}\delta(k^+-k^{\prime
+})\;.
\end{eqnarray}
The above expression can be further simplified into
Eq.~(\ref{z1d}) as $u^2\to 0$.

Following the reasoning for Eq.~(\ref{ad1}), the loop integral
associated with Fig.~1(b) is written as
\begin{eqnarray}
N^{(1)}_{b}&=&ig^2C_F \frac{2\pi i  u\cdot v}{u^-v^+-u^+v^-}
\int\frac{d^{2}l_T}{(2\pi)^{4}} \Bigg\{\frac{u^+\theta(k^{\prime
+}-k^+)\exp(-i{\bf l}_T\cdot {\bf b})} {(k^{\prime
+}-k^+)[2u^-(k^{\prime +}-k^+)^2+ u^+(l_T^2+m_g^2)]}
\nonumber\\
& &-\frac{v^+\theta(k^+-k^{\prime +})\exp(-i{\bf l}_T\cdot {\bf
b})} {(k^+-k^{\prime +})[2v^-(k^+-k^{\prime
+})^2+v^+(l_T^2+m_g^2)]}\Bigg\}\;, \label{ae1}
\end{eqnarray}
where the $\theta$-functions $\theta(k^{\prime +}-k^+)$ and
$\theta(k^+-k^{\prime +})$ correspond to the integration ranges
$l^+>0$ and $l^+<0$, respectively. The integration over $l_T$
gives
\begin{eqnarray}
N^{(1)}_{b}&=&-\frac{\alpha_sC_F}{\pi} \frac{u\cdot
v}{\sqrt{(u\cdot v)^2-u^2v^2}} \Bigg\{\frac{\theta(k^{\prime
+}-k^+)}{k^{\prime +}-k^+}
K_0\left(\sqrt{\frac{2u^-}{u^+}(k^{\prime +}-k^+)^2+m_g^2}b\right)
\nonumber\\
& &-\frac{\theta(k^+-k^{\prime +})}{k^+-k^{\prime +}}
K_0\left(\sqrt{\frac{2v^-}{v^+}(k^+-k^{\prime +})^2+m_g^2}b\right)
\Bigg\}\;. \label{ae2}
\end{eqnarray}
We split the above expression into
\begin{eqnarray}
N_b^{(1)}(k^+,k^{\prime +},b,\mu)=\delta(k^+-k^{\prime
+})\int_{0}^\infty dy
N_b^{(1)}(k^+,y,b,\mu)+N_{b+}^{(1)}(k^+,k^{\prime +},b,\mu)\;,
\label{ae3}
\end{eqnarray}
where the first term can be combined with $N^{(1)}_{a}$, and the
second term defines the ``plus" distribution. The first term is
rewritten, by applying the variable changes $y=(t+1)k^+$ for the
first Bessel function and $y=(1-t)k^+$ for the second Bessel
function, as
\begin{eqnarray}
\int_{0}^\infty dy N_b^{(1)}(k^+,y,b,\mu)
&=&-\frac{\alpha_sC_F}{\pi} \frac{u\cdot v}{\sqrt{(u\cdot
v)^2-u^2v^2}} \Bigg\{\int_0^\infty \frac{dt}{t^{1-2\delta}}
K_0\left(\sqrt{4\zeta^2t^2+m_g^2}b\right)
\nonumber\\
& &-\int_0^1 \frac{dt}{t^{1-2\delta}}
K_0\left(\sqrt{2k^{+2}t^2+m_g^2}b\right) \Bigg\}\;, \label{ae4}
\end{eqnarray}
where the denominators $t$ have been replaced by $t^{1-2\delta}$ as
in Eq.~(\ref{ad11}). In the asymptotic region with large $k^+$, it
is legitimate to extend the upper bound of $t$ in the second
integral from 1 to $\infty$. Using the relation
\begin{eqnarray}
\int_0^\infty x^{2\mu+1}K_0\left(\alpha\sqrt{x^2+z^2}\right)dx=
2^\mu\Gamma(\mu+1)\left(\frac{z}{\alpha}\right)^{\mu+1}K_{\mu+1}
(\alpha z)\;,
\label{ae5}
\end{eqnarray}
Eq.~(\ref{ae4}) becomes
\begin{eqnarray}
\int_{0}^\infty dy N_b^{(1)}(k^+,y,b,\mu)
=-\frac{\alpha_sC_F}{\pi} \frac{u\cdot v}{\sqrt{(u\cdot
v)^2-u^2v^2}} 2^{-1+\delta}\Gamma(\delta)K_\delta(m_gb)
\left[\left(\frac{m_g}{4\zeta^2 b}\right)^\delta
-\left(\frac{m_g}{2k^{+2} b}\right)^\delta \right]\;.\label{ae55}
\end{eqnarray}
Taking the limit $\delta\to 0$, the $1/\delta$ poles cancel
between the above two integrals, and Eq.~(\ref{ae55}) reduces to
\begin{eqnarray}
\int_{0}^\infty dy N_b^{(1)}(k^+,y,b,\mu)
=-\frac{\alpha_sC_F}{2\pi} \frac{u\cdot v}{\sqrt{(u\cdot
v)^2-u^2v^2}} \ln\frac{u^-v^+}{u^+v^-} \ln\frac{m_g
be^{\gamma_E}}{2}\;. \label{ae6}
\end{eqnarray}
The second term in Eq.~(\ref{ae3}) is written as
\begin{eqnarray}
N^{(1)}_{b+}(k^+,k^{\prime +},b,\mu)&=&-\frac{\alpha_sC_F}{\pi}
\frac{u\cdot v}{\sqrt{(u\cdot v)^2-u^2v^2}}\left[
\frac{\theta(k^{\prime +}-k^+)}{(k^{\prime +}-k^+)_+}
K_0\left(2\nu(k^{\prime +}-k^+)b\right)\right.
\nonumber\\
& &\left.-\frac{\theta(k^+-k^{\prime +})}{(k^+-k^{\prime +})_+}
K_0\left(\sqrt{2}(k^+-k^{\prime +})b\right)\right] \;, \label{ae7}
\end{eqnarray}
where the infrared regulators $m_g^2$ have been dropped, since a
plus distribution is infrared finite. The combination of
Eqs.~(\ref{ae6}) and (\ref{ae7}) then gives Eq.~(\ref{pde}).

The loop integral associated with Fig.~1(c) from
Eq.~(\ref{p2eBrv}) is rewritten, in the light-cone coordinates, as
\begin{eqnarray}
N^{(1)}_{c}&=&ig^2C_F\mu^{2\epsilon}
\int\frac{d^{4-2\epsilon}l}{(2\pi)^{4-2\epsilon}}
\frac{2(k^{\prime +}-l^+)u^-}{[2(l^+-k^{\prime +})l^--l_T^2]
(2l^+l^--l_T^2-m_g^2)(u^-l^++u^+l^-)}\delta(k^+-k^{\prime +})\;.
\label{p2e1}
\end{eqnarray}
For $0<l^+<k^{\prime +}$, we enclose the contour in the $l^-$
plane over the pole $l^-=(l_T^2+m_g^2)/(2l^+)-i\epsilon$. For
$k^{\prime +}<l^+$, we enclose the contour over the pole
$l^-=-u^-l^+/u^++i\epsilon$. For $l^+<0$, there is no pinch
singularity (noticing $u^+<0$ in our choice), and the contour
integration vanishes. Hence, Eq.~(\ref{p2e1}) becomes
\begin{eqnarray}
N^{(1)}_{c}&=&ig^2C_F\mu^{2\epsilon}4\pi
i\int\frac{d^{2-2\epsilon}l_T}{(2\pi)^{4-2\epsilon}}
\Bigg\{\int_0^{k^{\prime +}}dl^+ \frac{(k^{\prime
+}-l^+)l^+u^-}{[k^{\prime +}l_T^2+(k^{\prime +}-l^+)m_g^2]
[u^+(l_T^2+m_g^2)+2u^-l^{+2}]} \nonumber\\
& & -\int_{k^{\prime +}}^{\infty}dl^+ \frac{(l^+-k^{\prime
+})u^+u^-}{[u^+l_T^2+2u^-l^+(l^+-k^{\prime +})]
[u^+(l_T^2+m_g^2)+2u^-l^{+2}]}\Bigg\}\delta(k^+-k^{\prime +})\;.
\label{pe2}
\end{eqnarray}
The integration over $l_T$ leads to
\begin{eqnarray}
N^{(1)}_{c}&=&-\frac{\alpha_sC_F}{2\pi}
\left(\frac{4\pi\mu^2}{m_g^2}\right)^{\epsilon}
\Gamma(\epsilon)\Bigg\{\int_0^1dt t^{-1+2\delta}(1-t)^{-\epsilon}
-\int_0^\infty dt t^{-1+2\delta}\left(
\frac{4\zeta^2}{m_g^2}t^{2}+1\right)^{-\epsilon}
\nonumber\\
& &+\left(\frac{4\zeta^2}{m_g^2}\right)^{-\epsilon}
\left[\int_0^{\infty}dt t^{-2\epsilon}
-\int_1^{\infty}dtt^{-\epsilon}(t-1)^{-\epsilon}\right]
\nonumber\\
& &+\left(\frac{4\zeta^2}{m_g^2}\right)^{-\epsilon}
\int_1^{\infty}dtt^{-1-\epsilon}(t-1)^{-\epsilon}
-\int_0^1dt(1-t)^{-\epsilon}\Bigg\}\delta(k^+-k^{\prime +})\;,
\label{pe24}
\end{eqnarray}
where the variable change $l^+=k^{\prime +}t$ has been made. The
above expression has been arranged in a way that the infrared
divergences from $t\to 0$ cancel in the first line, and the linear
ultraviolet divergences cancel in the second line. We have
\begin{eqnarray}
N^{(1)}_{c}&=&-\frac{\alpha_sC_F}{2\pi}
\left(\frac{4\pi\mu^2}{m_g^2}\right)^{\epsilon}
\Gamma(\epsilon)\Bigg\{B(2\delta,1-\epsilon)-
\frac{1}{2}\left(\frac{4\zeta^2}{m_g^2}\right)^{-\delta}
B(\delta,\epsilon-\delta)
\nonumber\\
& &+\left(\frac{4\zeta^2}{m_g^2}\right)^{-\epsilon}
\frac{\epsilon}{1-2\epsilon}B(1-\epsilon,2\epsilon)
+\left(\frac{4\zeta^2}{m_g^2}\right)^{-\epsilon}
B(1-\epsilon,2\epsilon)
-\frac{1}{1-\epsilon}\Bigg\}\delta(k^+-k^{\prime +})\;,
\label{pe25}
\end{eqnarray}
which leads to Eq.~(\ref{peB1v}) by employing the expansion,
\begin{eqnarray}
\Gamma(\epsilon)\approx \frac{1}{\epsilon}\left[1
-\gamma_E\epsilon+\left(\frac{\gamma_E^2}{2}
+\frac{\pi^2}{12}\right)\epsilon^2\right]\;.
\end{eqnarray}

We calculate the correction from Fig.~1(d) in the light-cone
coordinates,
\begin{eqnarray}
N^{(1)}_{d}&=& -ig^2C_F\int\frac{d^4l}{(2\pi)^4} \frac{2(k^{\prime
+}-l^+)u^-}{[2(l^+-k^{\prime +})l^--l_T^2]
(2l^+l^--l_T^2-m_g^2)(u^-l^++u^+l^-)}
\nonumber\\
& &\times \delta\left(k^+-k^{\prime +}+l^+\right) \exp(-i{\bf
l}_T\cdot {\bf b})\;. \label{p2e10}
\end{eqnarray}
Applying the reasoning for Eq.~(\ref{p2e1}), we have
\begin{eqnarray}
N^{(1)}_{d}&=&-ig^2C_F4\pi i\theta(k^{\prime
+}-k^+)\int\frac{d^{2}l_T}{(2\pi)^{4}} \Bigg\{ \frac{k^+(k^{\prime
+}-k^+)u^-\exp(-i{\bf l}_T\cdot {\bf b})}{[k^{\prime
+}l_T^2+k^+m_g^2] [u^+(l_T^2+m_g^2)+2u^-(k^{\prime
+}-k^+)^{2}]} \nonumber\\
& & + \frac{k^+u^+u^-\exp(-i{\bf l}_T\cdot {\bf
b})}{[u^+l_T^2-2u^-k^+(k^{\prime +}-k^+)]
[u^+(l_T^2+m_g^2)+2u^-(k^{\prime +}-k^+)^{2}]}\Bigg\}\;.
\label{p2e11}
\end{eqnarray}
The integration over $l_T$ gives
\begin{eqnarray}
N^{(1)}_{d}&=&\frac{\alpha_sC_F}{\pi} \frac{k^+\theta(k^{\prime
+}-k^+)}{k^{\prime +}(k^{\prime
+}-k^+)}\left[K_0\left(\sqrt{k^+/k^{\prime +}}m_gb\right)
-K_0\left(\sqrt{\frac{2u^-}{u^+}(k^{\prime
+}-k^+)^2+m_g^2}b\right)\right]\;. \label{p2e12}
\end{eqnarray}
We then adopt the splitting similar to Eq.~(\ref{ae3}),
whose first term is rewritten, by applying the variable change
$y=k^+/(1-t)$ for the first Bessel function and $y=(t+1)k^+$ for
the second Bessel function, as
\begin{eqnarray}
\int_{0}^\infty dy N_d^{(1)}(k^+,y,b,\mu)
&=&\frac{\alpha_sC_F}{\pi}\left[\int_0^1 dtt^{-1+2\delta}
K_0\left(\sqrt{1-t}m_gb\right)-\int_0^\infty dt t^{-1+2\delta}
K_0\left(\sqrt{4\zeta^2 t^2+m_g^2}b\right)\right.
\nonumber\\
& &\left.+\int_0^\infty \frac{dt}{t+1} K_0\left(2\zeta
bt\right)\right]\;. \label{p2e131}
\end{eqnarray}
It is easy to show, using the relation,
\begin{eqnarray}
\int_0^\infty\frac{K_\nu(\alpha x)}{x+1}dx=
\frac{\pi^2}{2}\csc^2(\nu\pi)\left[I_\nu(\alpha)+I_{-\nu}(\alpha)
-e^{-i\nu\pi/2}{\bf J}_\nu(i\alpha)
-e^{i\nu\pi/2}{\bf J}_{-\nu}(i\alpha)\right]\;,
\end{eqnarray}
where ${\bf J}_0$ denotes the Anger function, that the third term
is given by $\pi/(4\zeta b)$ in the asymptotic region with large
$k^+$. Employing Eq.~(\ref{ae5}), Eq.~(\ref{p2e131}) becomes
\begin{eqnarray}
\int_{0}^\infty dy N_d^{(1)}(k^+,y,b,\mu)
&=&-\frac{\alpha_sC_F}{2\pi}\left[ \ln\frac{2\zeta^2 b
e^{\gamma_E}}{m_g}\ln\frac{m_gbe^{\gamma_E}}{2}
+\frac{\pi^2}{6}\right]\;. \label{p2e132}
\end{eqnarray}
The second term in the splitting with the plus distribution can be
obtained in a way similar to Eq.~(\ref{ae7}).


The $O(\alpha_s)$ correction from Fig.~1(e) is written as
\begin{eqnarray}
N^{(1)}_{e}&=&\frac{i}{4}g^2C_F \int\frac{d^4l}{(2\pi)^4}tr\left[
\frac{\gamma^\nu(\not k'-\not l)}{(k'-l)^2 (l^2-m_g^2)}
\gamma_5\not n_-\not n_+\gamma_5\right]\frac{v_\nu}{v\cdot l}
\nonumber\\
& &\times \delta\left(k^+-k^{\prime +}+l^+\right) \exp(-i{\bf
l}_T\cdot {\bf b})\;, \nonumber\\
&=&ig^2C_F\int\frac{d^4l}{(2\pi)^4} \frac{2(k^{\prime
+}-l^+)v^-}{[2(l^+-k^{\prime +})l^--l_T^2]
(2l^+l^--l_T^2-m_g^2)(v^-l^++v^+l^-)}
\nonumber\\
& &\times \delta\left(k^+-k^{\prime +}+l^+\right) \exp(-i{\bf
l}_T\cdot {\bf b})\;. \label{p2bB2}
\end{eqnarray}
The substitution of $v$ for $u$ introduces an
additional pole $l^-=v^-l^+/v^+-i\epsilon$ in the region $l^+ <
k^{\prime +}$. It is straightforward to confirm that the
contribution from this additional pole vanishes. Hence, the result
of Fig.~1(e) is similar to that of Fig.~1(d), but with the
variable $\zeta$ replaced by $k\cdot v/\sqrt{v^2}$, which is
Eq.~(\ref{p2bB1}).

\newpage

\begin{center}
\begin{picture}(250,450)(-50,-150)

\Line(-100,250)(-30,250) \Line(-100,253)(-30,253)
\Line(-100,185)(-30,185) \GlueArc(-30,250)(30,-180,-90){5}{4}
\Line(-30,220)(-30,250) \Line(-27,220)(-27,250)
\Text(-60,220)[]{$l$} \Text(-90,240)[]{$b$}
\Text(-60,160)[]{$(a)$}

\Line(0,250)(70,250) \Line(0,253)(70,253) \Line(0,185)(70,185)
\Gluon(40,250)(70,215){5}{4} \Line(70,185)(70,215)
\Line(73,185)(73,215) \Text(40,160)[]{$(b)$}

\Line(100,250)(170,250) \Line(100,253)(170,253)
\Line(100,185)(170,185) \Gluon(140,185)(170,220){5}{4}
\Line(170,220)(170,250) \Line(173,220)(173,250)
\Text(140,160)[]{$(c)$}

\Line(200,250)(270,250) \Line(200,253)(270,253)
\Line(200,185)(270,185) \GlueArc(270,185)(30,90,180){5}{4}
\Line(270,185)(270,215) \Line(273,185)(273,215)
\Text(240,160)[]{$(d)$}

\Line(-100,100)(-30,100) \Line(-100,103)(-30,103)
\Line(-100,35)(-30,35) \Gluon(-60,100)(-60,35){5}{4}
\Text(-60,10)[]{$(e)$}

\Line(0,100)(70,100)
\Line(0,35)(70,35) \GlueArc(35,100)(20,-180,0){5}{5}
\Text(40,10)[]{$(f)$}

\Line(100,100)(170,100)
\Line(100,35)(170,35) \GlueArc(135,35)(20,0,180){5}{5}
\Text(140,10)[]{$(g)$}



\end{picture}

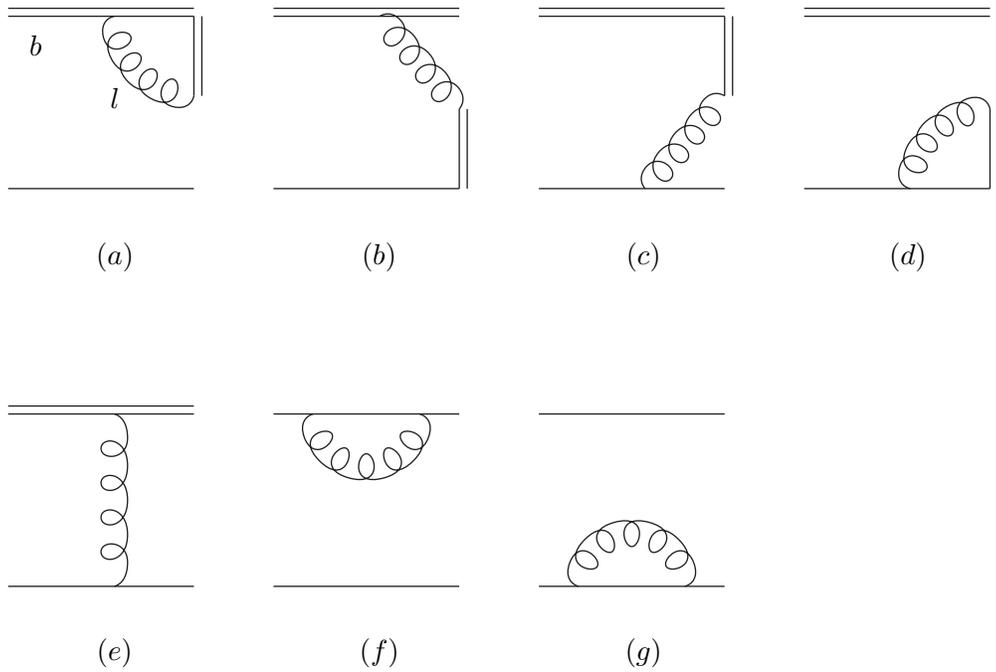
{\bf FIG. 1} $O(\alpha_s)$ diagrams for the $B$ meson wave
function.

\end{center}

\end{document}